\documentstyle[aps,multicol]{revtex}
\newcommand{\bqa}{\begin{eqnarray}}
\newcommand{\eqa}{\end{eqnarray}}
\newcommand{\beq}{\begin{equation}}
\newcommand{\eeq}{\end{equation}}

\begin{document}
\title{Evaluating the partition function for systems with long
range interactions}
\author{Tong Zhou}
\address{The James Franck Institute,
The University of Chicago, 
5640 S. Ellis Avenue,  
Chicago, IL 60637}
\date{\today}
\maketitle

\begin{abstract}

We express the partition function for an equilibrium system
of interacting particles in the canonical ensemble
as a functional integration over the particles'
density field.  We outline a method to evaluate
the partition function by expanding around a saddle
point.  The saddle point is shown to be the solution of the 
equivalent mean-field theory.
Leading corrections to the mean-field theory 
takes the form of a Gaussian integral.

\vspace{0.1in}
\end{abstract}
\begin{multicols}{2}

The partition function for a system of interacting particles in equilibrium
can be written down quite easily.  Using it to evaluate physically 
relevant quantities, however, generally requires aggressive approximations.
For example, expansion in powers
of the density is only good for short-range interactions and low density\cite{Landau}.
Few approximation schemes are appropriate for strongly interacting
systems characterized by long-ranged forces, such as Coulomb interactions and 
gravitational forces.

The method most often used to address systems with long range interactions
is the mean-field theory, in which, for any chosen particle,
the effects of all other particles are replaced by a mean-field
which is not affected by this particle.  Its advantage lies in its 
simplicity and physical clarity.  Due to the long-range
nature of the interactions, mean-field theory is often a good
approximation.  However, its inadequacy is shown in 
experiments\cite{LG,CG,Neu}.  

The lack of thermodynamical self-consistency of the
mean-field theory was pointed out nearly
seventy years ago\cite{Onsager,Kirkwood}.  
Efforts were then made to develop rigorous approaches for 
treating such systems by Onsager\cite{Onsager}, 
Kirkwood\cite{Kirkwood}, 
and Bogolyubov\cite{Landau}. 
These methods, from different mathematical view points,
however, all relied on detailed descriptions of the microscopic
states.  Attempts to treat such systems by finding descriptions
involving increasing number of particles are undermined by the
notorious closure problem.
Consequently these methods failed to make any practical
predictions, nor an estimation for the accuracy of the mean-field
theory.

Philosophically, the failure of these methods is due to the fact
that a system with long range interactions is highly coupled---any
part of it is affected by the changes in the whole system---and
cannot be described effectively by schemes focusing on local
effects.  This suggests an approach which treats such a system
as a whole.  The relevancy of the system as a whole also determines
that the detailed local information becomes more and more irrelevant
with increasing system sizes, and thus
can be replaced for a big system through coarse-graining.  

This Letter proposes
a new approach based on these considerations.
We express the partition function
as a functional integration over the particles' density field.  The saddle point
of this expression turns out to be equivalent to the mean-field theory.  
Corrections to the mean-field theory
then can be expressed as an expansion around this saddle point.

Let us consider a system with $N$ classical particles, volume $V$ 
and temperature $T$.  Its partition function in the canonical 
ensemble is,
\bqa
Z & = & \frac{1}{h^{3N}N!} \int_V e^{-\beta E} d\Gamma\\
& = & \frac{1}{N!}\left(\frac{2\pi mkT}{h^2}\right)^{\frac{3N}{2}}\int_V e^{-\beta U}
d\vec{q}_1\cdots d\vec{q}_N.
\eqa
where $U(\vec{q}_1\cdots \vec{q}_N)$ is the potential energy due to external fields or interactions among
particles.  In the following, we will concentrate on the integral part in the above
expression, so let us write,
\beq
Z_s=\int_V e^{-\beta U} d\vec{q}_1\cdots d\vec{q}_N.
\label{eq:zs}
\eeq

Equation (\ref{eq:zs}) is written in terms of integrations over the particles' positions.  
We want to instead express $Z_s$ in terms of functional integrations over
the particles' 
density field.  To derive such an expression, let us start with the ideal 
gas---the factor $\exp(-\beta U)$ can be simply inserted into the result later.  

For an ideal gas, $Z_s=V^N$.  Let us divide $V$ into $M$ cells, with volumes 
$V_1, V_2, \cdots , V_M$.  A state $\vec{n}$
 is represented by the number of particles
in these cells, $\vec{n}=(n_1, n_2, \cdots , n_M)$.  
The division must satisfy the 
condition that there are many particles in each cell, i.e.\ $n_i$ is big.
The probability weight of such
a state is denoted by $W(\vec{n})$ and it can be obtained 
through the following 
considerations.

\bqa
Z_s & = & (V_1+V_2+\cdots +V_M)^N\\
& = & \sum_{\vec{n}}^\prime \frac{N!}{n_1!n_2!\cdots n_M!}V_1^{n_1}V_2^{n_2}\cdots V_M^{n_M}\\
& \equiv & \sum_{\vec{n}}^\prime W(\vec{n}),
\label{eq:zw}
\eqa
where the $\prime$ above the summation sign indicates that particle number
is preserved: $\sum n_i =N$.

We can evaluate Eq.~(\ref{eq:zw}) using Stirling's approximation,
\beq
n_i! = \sqrt{2\pi n_i} n_i^{n_i} e^{-n_i},
\eeq
because $n_i$ is big.  Then
\bqa
W(\vec{n}) & = & \frac{N!}{\prod (\sqrt{2\pi n_i} n_i^{n_i}e^{-n_i})}
\prod_{i=1}^M V_i^{n_i}\\
& = & N! e^N \prod_{i=1}^M (V_i^{n_i}/\sqrt{2\pi n_i}) 
\exp\left[-\sum n_i\log n_i\right]
\eqa
And we express this in terms of the density $\rho_i=n_i/V_i$ of 
particles in cell $i$,
\beq
W(\vec{n})=N! e^N \prod_{i=1}^M (2\pi \rho_i V_i)^{-\frac{1}{2}} 
\exp\left[-\int \rho\log\rho dV\right].
\eeq
We use this form of $W(\vec{n})$ to evaluate $Z_s$,
and change from summations over different configurations of $n_i$ to 
integrations over continuous variables $n_i$,
\bqa
Z_s & = & N! e^N \sum_{\vec{n}}^\prime 
\prod_{i=1}^M (2\pi \rho_i V_i)^{-\frac{1}{2}}
\exp\left[-\int \rho\log\rho dV\right]\\
& = & N! e^N \int^\prime \prod_{i=1}^M 
\frac{d\rho_i}{\sqrt{2\pi \rho_i /V_i}}
\exp\left[-\int \rho\log\rho dV\right]\\
& \equiv & \sqrt{2 \pi N} N^N \int^\prime D\rho  
\exp\left[-\int \rho\log\rho dV\right].
\eqa

For general cases, we simply insert the
factor $e^{-\beta U}$, so
\beq
Z_s= \sqrt{2 \pi N} N^N \int^\prime D\rho e^{-\beta f},
\label{eq:masw}
\eeq
where 
\beq
\int^\prime D\rho \equiv \int^\prime 
\prod_{i=1}^M \frac{d\rho_i}{\sqrt{2\pi \rho_i /V_i}},
\eeq
and
\beq
f\equiv U +kT\int \rho\log\rho dV.
\label{eq:deff}
\eeq
Equations (\ref{eq:masw}-\ref{eq:deff}) are the desired expression
for $Z_s$ and are the central result of this letter.

When we replace the position variables of the particles with the 
coarse-grained density distribution, certain information is lost.
In particular, $U$ is calculated not for the actual 
positions of the particles, but instead for the coarse-grained density
distribution. For this to be a good approximation, 
certain quantities such as field potential must
not vary appreciably within a cell; each cell is treated as a
uniform subsystem.  This is an additional condition to the one
mentioned above that there are many particles in each cell.  
Finding an appropriate division of a macroscopic system which
satisfies both conditions only is possible if 
the range of the interaction is long enough.

To calculate $Z_s$ by using Eq.~(\ref{eq:masw}) directly is as difficult 
as by using Eq.~(\ref{eq:zs}).  However, the form of Eq.~(\ref{eq:masw})
suggests a different approach---first, find the saddle point, i.e.\ the
minimum of $f$, and then expand around this saddle point.

Let us assume $\rho_0$ is the density distribution which minimizes
$f$.  The corresponding minimum is $f_0$.  Then,
\beq
\left. {\frac{\delta f}{\delta \rho}}\right|_{\rho=\rho_0}=0.
\label{eq:saddle}
\eeq

To study the saddle point, let us consider an example of an 
system of particles
with Coulomb interaction, encompassed by a surface $\Sigma$
with surface charge $\sigma$. 
For simplicity, let us assume all particles carry charge $e$.
Then the interaction potential energy for a configuration $\rho$
of particles is
\beq
U=\frac{1}{2}\int\rho e \phi dV
+\frac{1}{2}\int_{\Sigma}\sigma \phi dS,
\eeq
where $\phi$ is the electric field potential, and is connected to $\rho$
by the Poisson equation,
\beq
\nabla^2\phi=-\frac{e\rho}{\epsilon_0},
\eeq
and to $\sigma$ at the boundary $\Sigma$ by,
\beq
\left(\frac{\partial \phi}{\partial n}\right)_{in}
-\left(\frac{\partial \phi}{\partial n}\right)_{out}
=\frac{\sigma}{\epsilon_0},
\eeq
where $\vec{n}$ is the unit vector normal to the surface, and pointing
outside the system. So, keeping in mind that $\int\delta\rho dV=0$,
\bqa
\delta f & = & \int dV\left[\frac{e}{2}(\rho_0\delta\phi+
\phi_0\delta\rho)+kT(1+\log\rho_0)\delta\rho\right]\nonumber\\
&&+\int_{\Sigma}\frac{\sigma \delta\phi}{2}dS \\
& = & \int dV (e\phi_0 + kT\log\rho_0)\delta\rho
+\int_{\Sigma} dS\left[\frac{\sigma\delta\phi}{2}\right.\nonumber\\
&&+\left.
\frac{\epsilon_0}{2}\phi_0\delta\left(\frac{\partial\phi}{\partial n}
\right)_{in}-\frac{\epsilon_0}{2}\left(\frac{\partial\phi_0}{\partial n}
\right)_{in}\delta\phi\right]\\
&=&\int dV (e\phi_0 + kT\log\rho_0)\delta\rho\nonumber\\
&&+\frac{\epsilon_0}{2}\int_{\Sigma}dS
\left[\phi_0\delta\left(\frac{\partial\phi}{\partial n}
\right)_{out}-\left(\frac{\partial\phi_0}{\partial n}
\right)_{out}\delta\phi\right].
\eqa
The second term in the last expression vanishes because there is
no charges outside of the system.  Then the vanishing of the first term
leads to the Boltzmann distribution,
\beq
\rho_0 =\varrho_0 e^{-\frac{e\phi_0}{kT}},
\label{eq:boltz}
\eeq
where the constant $\varrho_0$ ensures $\int\rho_0dV=N$.
Now, moreover, we find that the saddle point corresponds to the 
Poisson-Boltzmann (PB) equation,
\beq
\nabla^2\phi_0=-\frac{e\varrho_0}{\epsilon_0}e^{-\frac{e\phi_0}{kT}}.
\eeq

Even though the above example focused on the special case
of Coulomb interactions, Eq.~(\ref{eq:deff}) is completely general
so that the saddle point always correspond to the system's 
mean-field theory.  This is reasonable because physically, 
when we consider only the optimal density
distribution and ignore all fluctuations, 
we get the mean-field theory.  Of course, the mean-field theory, though
often a good approximation, is neither complete, nor thermodynamically 
self-consistent.  In particular it ignores fluctuations whose effects we will
consider next.

We account for fluctuations by
expanding around the saddle point, $\rho=\rho_0+\delta\rho$.  
The first-order change in $f$ vanishes, let us write the second order
change in $f$ as $\delta^2f$, and in $U$ as $\delta^2U$.
$\delta^2f$ is a quadratic functional of $\delta\rho$,
\beq
\delta^2f=\delta^2U+kT\int\frac{(\delta\rho)^2}{2\rho_0}dV,
\eeq
and thus is always positive.  
For Coulomb and gravitational interactions, there are no
higher order corrections to $U$ than $\delta^2U$.

Now we have, from Eq.~(\ref{eq:masw}),
\beq
Z_s=\sqrt{2\pi N} N^N e^{-\beta f_0} 
\int^{\prime}D\rho e^{-\beta \delta^2f},
\label{eq:zscon}
\eeq
where
\beq
\int^\prime D\rho \equiv \int^\prime 
\prod_{i=1}^M \frac{d\rho_i}{\sqrt{2\pi \rho_{0i} /V_i}}.
\label{eq:nD}
\eeq
In obtaining Eq.~(\ref{eq:zscon}), we neglect terms to the order of
$(\delta\rho)^3$ in comparison to terms to the order of
$(\delta\rho)^2$ in the expansion of $\rho\log\rho$, also in
(\ref{eq:nD}), we write $\rho_{0i}$ instead of $\rho_i$ in the
denominator.  Both approximations
are justified by the fact that each cell
contains many particles.
The prime above the integration sign in Eq.~(\ref{eq:zscon})
indicates that particle number is preserved, $\int\rho dV=N$.  

Now we can write,
\beq
Z_s=Z_0Z_m,
\label{eq:Z}
\eeq
where
\beq
Z_0=N^Ne^{-\beta f_0},
\eeq
and
\beq
Z_m=\sqrt{2\pi N} \int^{\prime}D\rho e^{-\beta \delta^2f}.
\label{eq:master}
\eeq

Thus, we separate $Z_s$ into two parts: $Z_0$ describes the mean-field 
theory, while $Z_m$ accounts for fluctuations.
For non-interacting particles, $\delta^2U=0$ and the Gaussian integral
for $Z_m$ gives $Z_m=1$.  The mean field theory is exact for this case, 
as it should be.  However, for interacting particles, 
$Z_m$ yields a nontrivial modification.

For example, ref.\cite{CGZ} investigates a model colloidal system of a 
charged sphere surrounded by counter-ions inside a spherical shell,
mimicking the Wigner-Seitz cell in a colloidal crystal.  The
mean-field (PB) theory predicts that the free energy is a monotonicly decreasing
function of the radius of the shell.  However, $Z_m$ contributes a increasing 
component to the free energy thus leads to 
a local minimum in the free energy's dependence on the shell's radius.
This would imply an effective attraction between like-charge colloidal particles,
which was observed in experiments\cite{LG,CG}.

If, instead of charged ions, we have neutral particles which interact with attractive
gravitational forces in the above model,  the mathematical expressions would
be nearly identical except that the interaction is with a different sign.  Then
immediately we see $Z_m$ would lead to an extra effective {\it repulsion}.
This may be relevant to the observation of anomalous gravitational
effects\cite{fifth}.

Let us consider the effects of dividing the system into cells from a different
view point.  In order to derive 
Equations~(\ref{eq:Z}-\ref{eq:master}), 
we need the condition that
in each cell there are many particles.  Of course, the final results should
not depend on the way how we divide the system.  Essentially, we are 
considering the effects of density fluctuations.  If we decompose these fluctuations
into modes with different wavelengths, then the size of the cell corresponds to
a short-wavelength cutoff.  When we calculate Eq.~(\ref{eq:master}), there are two
contributions: one from $\delta^2 U$ and the other from the integration of 
$\frac{(\delta\rho)^2}{2\rho_0}$.  The latter is totally 
insensitive to the way the cells are made---if we can ignore $\delta^2U$, we
always get $Z_m=1$ no matter how we divide the system.  
For long range interactions, we would indeed
expect $\delta^2U$ to be very insensitive to short wavelength fluctuations.
So the final result should be independent of the way the system is 
divided.

Equations (\ref{eq:Z}-\ref{eq:master}) remind us of 
a basic concept:  The density 
distribution is not one of the macroscopic quantities such as temperature
and volume which determine the equilibrium state of the system.  It cannot
be preset, but rather  results from an 
ensemble average.  Thus when there are interactions between particles,
thermodynamic properties cannot be calculated from a {\it fixed} optimal
density distribution in a self-consistent way.  The formalism developed
here remedies this deficiency.

The method developed here bears a passing resemblance 
to the density functional 
theory (DFT)\cite{KV,AM,Hansen,Davis}
in that both approximations use density as 
the basic variable.  However, DFT 
calculations typically are based on an optimal density distribution,
since the goal is often to calculate a system's ground state.
Here, density fluctuations are a essential component
as they lead to Eq.~(\ref{eq:master}).
For finite temperature, DFT calculations based on one optimal density
distribution cannot be thermodynamically self-consistent for reasons
stated above.  In that case, the method developed here may be 
helpful to extend DFT calculations.

Also, Ornstein derived the van der Waals equation in a very similar
way\cite{Ornstein}.  But again, the purpose there was to find the optimal
density distribution and no fluctuations were considered.

Netz and Orland recently
formulated a field theoretical approach\cite{NO} to the 
same class of problems addressed herein.  
They obtained the PB equation as a saddle point and 
calculated corrections by an diagrammatic expansion.  This approach is well
developed in field theory.  However, the convergence of the expansion
relies on a small parameter which is usually missing when the interaction
is long ranged.  Actually, Coalson and Duncan had shown that the subsequent 
terms in such an expansion are of the same order\cite{Caolson}.
The method developed here, on the other hand,  
exploits the long range of the interactions to coarse grain the system 
effectively.  

This work is motivated by the experimental results showing the inadequacy
of the mean-field theory\cite{LG,CG}.  For an application of the method
described here to a model colloidal system, see\cite{CGZ}.
The author wants to thank David Grier, Susan Coppersmith 
and Daniel Chung for helpful discussions.
This work was supported  in part by the  MRSEC Program of the NSF under award
number \#DMR-9400379.

\end{multicols}

\end{document}